\documentclass[aps,prx,twocolumn,floatfix,longbibliography,nofootinbib]{revtex4-2}
\usepackage[utf8]{inputenc}
\usepackage{natbib}
\usepackage{graphicx}
\usepackage{xcolor}
\usepackage[tbtags]{amsmath}
\usepackage[colorlinks=true, linkcolor=blue]{hyperref}
\usepackage{amssymb}
\usepackage{gensymb}
\usepackage{comment}

\usepackage{newpxmath}
\usepackage{float}
\usepackage{amsmath}
\usepackage{tabularx,graphicx}
\usepackage{epstopdf}
\usepackage{latexsym}
\usepackage{color, colortbl}
\usepackage{psfrag}
\usepackage{bbm}
\usepackage{bm}
\usepackage{titlesec}
\usepackage{dsfont}
\usepackage{feynmp}
\usepackage{slashed}
\usepackage{subcaption}
\usepackage{multirow}
\usepackage{subcaption}
\usepackage[newcommands]{ragged2e}
\usepackage[normalem]{ulem}

\usepackage{braket}
\usepackage{tcolorbox}
\usepackage{subcaption}
\usepackage{relsize}
\def \a {\alpha}

\def \b {\beta}
\def \im {\tn{Im}}

\def \B {\mathrm{B}}

\def \beq {\begin{eqnarray}}
\def \eeq {\end{eqnarray}}
\def \tn {\textnormal}

\def \M{{\cal {M}}}

\def \B {{\cal {B}}}

\def \nn {\nonumber}

\def \la{\langle}
\def \ra{\rangle}

\def \cj {\mathlarger{\chi}_{\scriptscriptstyle JJ}}
\def \sj {\mathlarger{S}_{\scriptscriptstyle J}}
\def \wi {\mathrm{W}_{\scriptscriptstyle JJ}}
\def \rs {\tn{Re}[\sigma_\beta(\omega)]}
\def \gw {\Gamma_\beta(\omega)}
\def \g {\Gamma_\beta}

\begin{document}
\title{Planckian Bounds via Spectral Moments of Optical Conductivity}
\author{Debanjan Chowdhury}\email{debanjanchowdhury@cornell.edu}
\affiliation{Department of Physics, Cornell University, Ithaca NY 14853, USA}

\begin{abstract}
The observation of Planckian scattering, often inferred from Drude fits in strongly correlated metals, raises the question of how to extract an intrinsic timescale from measurable quantities in a model-independent way.
We address this by focusing on a ratio ($\B$) of spectral moments of the dissipative part of the optical conductivity and prove a rigorous upper bound on $\B$ in terms of the Planckian rate. The bound emerges from the analytic structure of thermally weighted response functions of the current operator. Crucially, the bounded quantity is directly accessible via optical spectroscopy and computable from imaginary-time correlators in quantum Monte Carlo simulations, without any need for analytic continuation. We evaluate $\B$ for simplified examples of both Drude and non-Drude forms of the optical conductivity with a single scattering rate in various asymptotic regimes, and find that $\B$ lies far below the saturation value. These findings demonstrate that Planckian bounds can arise from fundamental constraints on equilibrium dynamics, pointing toward a possibly universal structure governing transport in correlated quantum matter.
\end{abstract}

\maketitle

{\it Introduction---} Is there a fundamental limit to how fast thermal many-body quantum systems equilibrate? Does this imply a bound on transport scattering rates? Interest in these questions has surged, driven by experiments across diverse quantum materials \cite{JANBruin2013,legrosUniversalTlinearResistivity2019a,YCao2020,Paglione,AP21,Grissonnanche2020} exhibiting ``high"-temperature superconductivity. In these systems, the transport scattering rate,
\beq
\Gamma_P = \frac{\a k_B T}{\hbar},
\eeq
depends only on temperature ($T=\beta^{-1}$), with an empirical prefactor $\a\sim O(1)$ \cite{JZaanen2004,SSachdev1999b}. Whether $\a$ admits a sharp universal bound remains a subject of considerable debate. This Planckian timescale is a unifying feature across microscopically distinct strongly correlated metals \cite{SAHartnoll2022}, contrasting sharply with Fermi liquids \cite{abrikosov2012methods}, where the {\it inelastic} rate scales as $\Gamma_{\rm{FL}}=g^2(k_B T)^2/W \ll \Gamma_P$, with $g$ a dimensionless interaction and $W$ a characteristic bandwidth. While a unified microscopic theory for Planckian scattering from electronic interactions in realistic lattice models remains elusive, progress has come from metallic quantum criticality \cite{SSLee2018,EBerg2019,SSachdev2023}, holographic dualities linking transport to chaotic black holes \cite{hartnoll2018holographic,PWPrev}, and exactly solvable models  \cite{DChowdhury2022a,Esterlis}. While insightful, these rely on assumptions that do not immediately extend to generic lattice Hamiltonians.

Extracting an intrinsic inelastic transport scattering rate from the dc resistivity is often nontrivial. Experimental estimates often use Drude-model fits \cite{JANBruin2013} with carrier densities and masses from independent low-temperature measurements; the systematic uncertainties can complicate extracting the precise value of $\a$. Even conventional metals display Planckian scattering at intermediate temperatures from well-understood {\it quasi-elastic} electron-phonon scattering in the equipartition regime \cite{ziman2001electrons}. Conjectured bounds typically apply to intrinsically {\it inelastic} processes; stability bounds likely forbid Planckian quasi-elastic rates with large coefficients \cite{CMurthy2023}. The absence of a sharp, model-independent definition of a ``universal" scattering rate directly related to measurable correlations has motivated alternative approaches, e.g., bounding charge diffusivity \cite{SAHartnoll2015,MBlake2016,ALucas2016,lucashartnoll17,hartnoll17,hartnoll18,Luca25}.

    Temperature-dependent optical conductivity, $\sigma_\beta(\omega)$, \cite{Basov_RMP} offers a more direct probe of a frequency-dependent scattering rate, $\gw$, from the dissipative response \cite{memory}. In numerous correlated metals, the low-frequency limit of $\gw$ shows Planckian scaling \cite{orenstein,Marel2003,basov07,YRS20,Michon2023,DChaudhuri2025,LiangWu}. However, we lack a model-independent theory for extracting a {\it possibly} temperature-bounded relaxation timescale from $\sigma_\beta(\omega)$. Here, we prove a rigorous bound on a specific frequency-weighted integral of optical conductivity, formulated entirely in terms of experimentally measurable quantities, that are also accessible to numerical methods, such as quantum Monte Carlo (QMC) \cite{becca2017quantum} without any further assumptions or analytic continuation. The Planckian time scale emerges naturally in the specific spectral moment ratio from analyticity and fundamental equilibrium constraints encoded in the KMS condition \cite{abrikosov2012methods}.

{\it Statement of the bound---} Our proposed bound is,
\begin{subequations}
\beq
&&\B \equiv \frac{\int_\omega g_\beta(\omega) ~\sj(\omega) }{\int_\omega \sj(\omega)} \leq \frac{C}{\beta^2},\label{bound1}~\tn{where} \\
&&g_\beta(\omega) = \frac{\omega^2}{[2\cosh(\beta\omega/2)]}. \label{gw}
\eeq
\end{subequations}
\noindent $C$ here is a dimensionless number that is independent of the microscopic Hamiltonian, whose numerical value follows purely from analytic properties of a complex function \cite{si}, introduced below. The current noise spectrum,
\begin{subequations}
\beq
\sj(t) &=& \frac{1}{2}\la \{ J(t), J\}\ra_\beta,\\
\sj(\omega) &=& \frac{\omega}{\tanh(\beta\omega/2)}\rs,
\eeq
\end{subequations}
where $J(t) = e^{iHt} J e^{-iHt}$ is the current operator and $H$ represents the many-body Hamiltonian. The current noise spectrum is related to the dissipative part of $\sigma_\beta(\omega)$, 
\begin{subequations}
\beq
\rs &=& \frac{\im~ \cj(\omega)}{\omega},\\
\cj(t) &=& -i\theta(t)\la [J(t),J]\ra_\b.\label{ret}
\eeq
\end{subequations} 
The expectation values are computed with respect to the thermal density matrix, $\la\cdots\ra \equiv \tn{Tr}[e^{-\beta H}\cdots]/Z$, where $Z=\tn{Tr}[e^{-\beta H}]$ is the partition function. 
 In what follows, we consider the longitudinal response along, say, the $x-$direction, $J \equiv J_x$. We will be using the shorthand, $\int_\omega\equiv \int_0^\infty d\omega$, throughout and work in units where $\hbar=1$. While $\B$ can depend on microscopic coupling constants and energy scales of $H$, it is bounded from above by a parameter-free universal Planckian rate. 

The frequency integrated noise spectrum in the denominator of $\B$ in Eq.~\eqref{bound1} measures average equilibrium current fluctuations. Thus $\B\equiv\la g_\beta(\omega)\ra$ is a moment w.r.t. the normalized distribution $P(\omega) = \sj(\omega)/\int_{\omega'} \sj(\omega')$. The function $g_\beta(\omega)$ has asymptotic forms: $g_\beta(\omega)\sim \omega^2
~~(\omega\ll\beta^{-1})$, $g_\beta(\omega)\sim e^{-\beta\omega/2}~~(\omega\gg \beta^{-1})$, picking out the characteristic frequency fluctuations centered around $\beta^{-1}$, with a width set by $\beta^{-1}$. The origin of the specific form of $g_\beta(\omega)$ in the bound will be explained below. 

{\it Connection to imaginary-time correlators ---} Before proving the bound in Eq.~\eqref{bound1}, we reformulate the statement using imaginary-time correlators, which are readily computable in QMC simulations. The imaginary-time current-current correlator is
\beq
\Lambda(\tau) = \la J(\tau) J(0)\ra,~~0<\tau<\beta,
\eeq
where $J(\tau)=e^{\tau H}Je^{-\tau H}$. This can be expressed in terms of the dissipative part of the conductivity as \cite{MR1,MR3,SLederer2017},
\beq
\Lambda(\tau) = \int_{-\infty}^\infty\frac{d\omega}{2\pi}\omega\rs\frac{\cosh\bigg[\bigg(\frac{\beta}{2}-\tau\bigg)\omega\bigg]}{\sinh(\beta\omega/2)}.\label{imcorr}
\eeq
This can be brought to a more familiar ``centered" form, $\widetilde\Lambda(\tau) \equiv \Lambda(\tau+\beta/2)$, where now $-\beta/2<\tau<\beta/2$, and
\beq
\widetilde\Lambda(\tau) = \int_{-\infty}^\infty\frac{d\omega}{2\pi}\omega\rs\frac{\cosh(\tau\omega)}{\sinh(\beta\omega/2)}.
\eeq
Eq.~\eqref{bound1} can then be written equivalently as \cite{si},
\beq
\B = \frac{\partial_\tau^2\Lambda(\tau=\beta/2)}{\Lambda(\tau=0)} \bigg[= \frac{\partial_\tau^2\widetilde\Lambda(\tau=0)}{\widetilde\Lambda(\tau=\beta/2)}\bigg] \leq \frac{C}{\beta^2},\label{imcurv}
\eeq
which states that the curvature of $\Lambda(\tau)$ at $\tau=\beta/2$ normalized by the value of $\Lambda(0)=\la J^2(0)\ra$ is bounded.

Several remarks are now in order. The correlator $\Lambda(\tau)$ satisfies $\Lambda(\tau) = \Lambda(\beta - \tau)$, so ``late-time" behavior corresponds to the longest accessible time $\tau = \beta/2$. As noted above, $\B$ is dominated by $|\omega| \lesssim T$, and thus late imaginary-time dynamics encodes moments of the response function over a narrow frequency range. QMC studies of a quantum critical metal have used the same imaginary-time correlators, including $\partial_\tau^2\Lambda(\tau = \beta/2)$, to construct a proxy for the dc resistivity at low$-T$, assuming that the low-frequency ($\omega \lesssim T$) conductivity is well described by a single Drude-like component with an $O(\beta^{-1})$ width \cite{SLederer2017}. Variants of such imaginary-time correlators for the Raman response have also been extracted directly from experimental data \cite{Lederer20}. However, the universal Planckian bounds on $\B$ regardless of any microscopic details being proposed here have not been identified in any previous works.

{\it Proof of Planckian bound ---} The bound follows from analyticity properties of the following Wightman function, obtained by symmetrically splitting the thermal Boltzmann factor,
\beq
\wi(t) = \frac{1}{Z}\tn{Tr}\bigg[e^{-\beta H/2} J(t) e^{-\beta H/2} J \bigg].\label{wightman}
\eeq
This can be expressed in the Lehmann representation as,
\beq
\wi(t) = \frac{1}{Z}\sum_{m,n}|J_{mn}|^2 e^{-\beta(E_m+E_n)/2} e^{i(E_n-E_m)t}, \label{wightman1}
\eeq
where $\{E_m,|m\ra \}$ represent the many-body energy eigenstates of $H$, with $J_{nm}=\la n|J|m\ra$. Since the conductivity can likewise be written as,
\beq
&&\rs = \nn\\
&&\frac{\pi}{\omega Z}\sum_{m,n} |J_{nm}|^2 (e^{-\beta E_m} - e^{-\beta E_n})\delta(\omega + E_m -E_n),\label{lehman}
\eeq
the two are related via \cite{DStanford2016a,DChowdhury2017}, 
\beq
\widetilde\wi(\omega) = \frac{\tn{Im}~\cj(\omega)}{2\sinh(\beta\omega/2)}   = \frac{\omega/2}{\sinh(\beta\omega/2)}~\rs. 
\eeq
For $\beta\omega\ll1$, $\widetilde\wi(\omega)\approx T~\rs$, while for $\beta\omega\gg1$, $\widetilde\wi(\omega)$ is exponentially suppressed in $\beta\omega$. Thus, $\widetilde\wi(\omega)$ naturally picks up contributions from the optical spectrum at frequencies smaller than the temperature, while suppressing the high-frequency part.

The Wightman function possesses useful analytic properties. Consider the analytic continuation $\wi(z)$ to a complex $z=t + i\tau$ in Eq.~\eqref{wightman1}. Up to oscillatory phases, the modulus of each term in the sum takes the form $ e^{-\beta(E_n+E_m)/2} e^{-(E_n - E_m)\tau}$. Convergence requires $-\beta/2 < \tau < \beta/2$, implying that $\wi(z)$ is analytic in the thermal strip $|\mathrm{Im}(z)| < \beta/2$. This is the symmetrized analog of the KMS strip associated with $G^>(z) = \mathrm{Tr}[e^{-\beta H} J(z) J(0)]/Z$, which is analytic for $0<\mathrm{Im}(z)<\beta$. Since $\wi(z)$ is not generically a constant function, it attains its maximum value at the boundary of the analytic domain, $|\wi(z)|\leq \wi(\pm i\beta/2)$, where,
\beq
\wi(\pm i\beta/2) = \frac{1}{Z}\sum_{m,n}|J_{mn}|^2 e^{-\beta E_m}.
\eeq
It is not a coincidence that the denominator of $\B$,
\beq
\int_\omega \sj(\omega) = \Lambda(\tau=0) = \wi(\pm i\beta/2).
\eeq
Let us construct the complex function,
\beq
F(z) \equiv \frac{\wi(z)}{\wi(i\beta/2)},
\eeq
where $\tn{Im}(z)=0$ corresponds to the real time axis as before. Clearly, $F(z)$ is analytic in the strip defined by $|\tn{Im}(z)|<\beta/2$, with $|F(z)|\leq 1$. From complex analysis, the second derivative for any real $t$ satisfies \cite{boas2011entire},
\beq
|F''(t)|\leq \frac{C}{\beta^2}, \label{gpp}
\eeq
where a conservative estimate leads to $C=8$, which can be further tightened to $C=16/\pi$ \cite{si}. 

Recall that the Wightman function, 
\beq
\wi(t) = \int_{-\infty}^\infty \frac{d\omega}{2\pi}~e^{-i\omega t}~\frac{\omega/2}{\sinh(\beta\omega/2)}~\rs,
\eeq
with $\B=-F''(t)|_{t=0}$, leading to Eq.~\eqref{bound1}. In summary, we proved that moments of the equilibrium current fluctuation, constructed from the Wightman function in the thermal strip, are bounded by a Planckian rate.

{\it Implications for ETH ansatz ---} Given the fundamental nature of the bound in Eq.~\eqref{bound1}, what structure does it impose on the current matrix elements $J_{mn}$ governing low-frequency conductivity? Let us parametrize the off-diagonal matrix elements using the eigenstate thermalization hypothesis (ETH) ansatz \cite{LDAlessio2016},
\beq
J_{mn} \simeq e^{-S(E)/2}f_J(E,\omega) R_{mn},
\eeq
where $E=(E_m+E_n)/2,~\omega=E_m-E_n$, $S(E)$ is the macrocanonical entropy with the density of states, $\rho(E)\sim e^{S(E)}$, and $R_{mn}$ are random variables with unit variance. Using the form of the conductivity in Eq.~\eqref{lehman}, and expanding $S(E\pm\omega/2)\approx S(E)\pm(\omega/2)S'(E)+...$ for small $\omega$, we find,
\begin{subequations}
\beq
&&\B = \frac{\int_\omega \la f^2_J(E,\omega)\ra_\beta~\omega^2}{\int_\omega \la f^2_J(E,\omega)\ra_\beta\cosh\bigg(\frac{\beta\omega}{2}\bigg)},~\tn{where}~\\
&&\la f^2_J(E,\omega)\ra_\beta = \frac{1}{Z}\int_E e^{S(E)-\beta E}|f_J(E,\omega)|^2.
\eeq
\end{subequations}
Assuming $f_J(E,\omega)$ varies slowly with $E$ over the thermal window, the $E-$integral can be evaluated by saddle point, fixing $E = E_\beta$ as the solution to $S'(E) = \beta$. This yields $\langle f_J(E,\omega)\rangle_\beta \approx f_J(E_\beta,\omega) \equiv f_J(\omega)$. The bound in Eq.~\eqref{bound1} then suggests,
\beq
\B = \frac{\int_\omega |f_J(\omega)|^2~\omega^2}{\int_\omega |f_J(\omega)|^2~\cosh(\beta\omega/2)}\leq \frac{C}{\beta^2}.
\eeq
The finiteness of $\mathrm{Tr}[\rho J^2]$ ensures that the high-frequency falloff satisfies at least $f_J(\omega) \sim e^{-\beta|\omega|/4}$ \cite{abanin}. The bound formulated in terms of $f_J(\omega)$ goes beyond the ETH ansatz by incorporating the analyticity and KMS conditions on the Wightman correlator. In the ETH language, $\B\equiv\la g_\beta(\omega)\ra$ is a moment w.r.t. the normalized distribution $\widetilde{P}(\omega)= |f_J(\omega)|^2\cosh(\beta\omega/2)/\int_{\omega'}|f_J(\omega')|^2\cosh(\beta\omega'/2)$, which must be peaked over an energy interval $\Delta E\sim \beta^{-1}$. 

{\it Illustrative examples ---} The discussion so far has been rather general. We now turn to specific forms of $\rs$ that arise in distinct physical settings and examine various asymptotic regimes of the bounded quantity, $\B$. 
We consider two concrete examples. The first is the standard Drude formula in the quasielastic limit \cite{Basov_RMP},
\beq
\sigma_\beta(\omega) = \frac{D}{\g - i\omega},\label{drude}
\eeq
where $D$ is the optical Drude weight, $\g$ is the optical scattering rate (ignoring any $\omega-$dependence), and we have ignored any dynamical mass renormalization, $m_\star(\omega)/m$ \cite{memory}. This limit describes e.g. impurity scattering (where $\g$ is temperature-independent) or weak-coupling electron-phonon scattering at high temperatures (where $\g\propto \beta^{-1}$). We also consider the Fermi-liquid regime, where $\gw$ exhibits both frequency and temperature dependence due to umklapp scattering. Note that the above functional form of $\sigma_\beta(\omega)$ is pathological at high frequencies and not meant to apply in a real system, where it has to fall off much faster. For our computations below, we will instead impose a finite UV cutoff on $\omega<W$ set by the electronic bandwidth. The second example is a Gaussian peak centered at $\omega = 0$, 
\beq
\rs=\frac{D}{\g}e^{-(\omega/\g)^2},\label{gaussian}
\eeq
which arises in various strong-coupling settings at intermediate to high temperatures, where the energy differences contributing to $\rs$ are quasi-random \cite{lindner10,mousatov19,JFMV21}. A natural question is whether the bound in Eq.~\eqref{bound1} constrains $\g$, and if these $\omega-$dependent forms of $\rs$ come close to saturating the bound.

We begin with Eq.~\eqref{drude}, where the frequency moments can be evaluated analytically \cite{si}. We focus on two asymptotic regimes. First, consider the high-temperature regime where $\g\ll\beta^{-1}$, where $\int_\omega \sj(\omega)\approx D/\beta$ and $\int_\omega g_\beta(\omega)\sj(\omega)\approx D\g/\beta^2$,
\beq
\B\stackrel{\beta\g\rightarrow0}{\mathlarger{\approx}} \bigg(\frac{\beta\g}{\pi} \bigg)~\frac{\pi^2}{\beta^2}.
\eeq
We find that the bounded quantity is limited by the narrow width of $\rs$ and lies well below saturation. Notice that the scaling is not a naive $\g^2$ despite the peak width $\g \ll T$; the tail of $\rs\sim 1/\omega^2$ receives appreciable weight from higher frequencies, shifting the scaling from $\g^2$ to $\g T$.
On the other hand, in the low-temperature regime where $\beta^{-1}\ll\g\ll W$, we find $\int_\omega\sj(\omega)\approx D\g\ln(W/\g)$ and $\int_\omega g_\beta(\omega)\sj(\omega)\approx D/\beta^4\g$,
\beq
\B \stackrel{\beta\g\gg 1}{\mathlarger{\approx}} \bigg(\frac{1}{(\beta\g)^2\ln(W/\g)} \bigg)\frac{1}{\beta^2}. \label{wide1}
\eeq
We find once again that the bounded quantity lies well below saturation. 

Next we consider the Fermi-liquid limit of Eq.~\eqref{drude}, with $\gw = (\omega^2 + 4\pi^2 T^2)/W$, where $W$ is the bandwidth as introduced previously. In the low-temperature regime, $\beta W\gg1$, we find $\int_\omega \sj(\omega)\approx W$ and $\int_\omega g_\beta(\omega)\sj(\omega)\approx1/W\beta^4$,
\beq
\B \stackrel{\beta W\gg1}{\mathlarger{\approx}} \bigg(\frac{1}{\beta W}\bigg)^2~\frac{1}{\beta^2},
\eeq
which lies far below the saturation value, as expected.

Finally, we turn to the non-Drude conductivity in Eq.~\eqref{gaussian}, focusing on two asymptotic limits. In the narrow-Gaussian regime $\beta\g \ll 1$, we find $\B \sim \g^2$, consistent with our naive expectations and far below the bound. In the opposite limit of a broad Gaussian, $\beta\g \gg 1$, the asymptotic form is,
\beq
\B\stackrel{\beta\g\gg 1}{\mathlarger{\approx}}\bigg(\frac{1}{\beta\g} \bigg)^2\frac{1}{\beta^2}, \label{wide2}
\eeq
which is reminiscent of Eq.~\eqref{wide1} (without the $\log$). Thus, the bounded quantity lies well below saturation, as is the case for all other examples considered here, even in the limit of a large scattering rate, $\g\gg\beta^{-1}$. At least for the explicit examples of $\rs$ considered here, the bound in Eq.~\eqref{bound1} does {\it not} immediately translate into a bound on $\g$.  

{\it Outlook --- } We have established that the analytic properties of equilibrium current correlation functions, related to the Wightman functions via additional thermal factors, strongly constrain their frequency moments by a Planckian rate. The mathematical structure underlying our derivation is reminiscent of bounds on the many-body Lyapunov exponent from out-of-time-ordered correlators (OTOCs) \cite{JMaldacena2016a}. However, unlike OTOC-based bounds, the bounds established here are experimentally testable in solid-state systems. They can be probed directly via optical measurements, and via imaginary-time correlators computed using QMC for strongly correlated lattice models. This opens a new avenue for uncovering universal structure in complex charge transport. A new perspective on the bounds presented here based on Lanczos coefficients might also be useful \cite{auerbach18,DEParker2019}.

This raises a natural question regarding connection to experiment. Recent work has performed a careful analysis of optical conductivity data in La$_{2-x}$Sr$_x$CuO$_4$ at $x=0.24$, finding excellent agreement with a generalized Drude form exhibiting $\omega/T$ scaling over a broad range of frequencies and temperatures \cite{Michon2023}. Specifically, the data are well described by the scaling ansatz $\gw = \beta^\zeta A_\Gamma(\beta\omega),~\frac{m_\star(\omega)}{m} = \beta^{\zeta+1}A_m(\beta\omega)$, where $\zeta$ is a power-law exponent and $A_\Gamma$, $A_m$ are scaling functions of $\beta\omega$. Numerical evaluation of how closely this scaling form approaches saturation would be a natural next step. Extending this analysis to other recent experiments may also reveal broader patterns \cite{YRS20,DChaudhuri2025,LiangWu}. Similarly, it would be worthwhile to revisit numerical results for imaginary-time correlators in quantum critical metals \cite{EBerg2019} and examine how the moments evolve as a function of tuning parameters across the many-body phase diagram, as a way to potentially extract the relevant intrinsic relaxation timescales.

Much of the discussion surrounding Planckian scattering has focused on identifying mechanisms that yield a $T-$linear scattering rate. Our results point instead toward a deeper question. To what extent do fundamental constraints on thermal quantum matter, as encoded in the analytic structure of correlation functions, dictate their dynamical behavior, including transport? The emergence of Planckian scales from such general principles, suggests that universal structure governing relaxation in interacting systems may be far more robust than previously appreciated. Uncovering further bounds of this kind, amenable to both experimental measurement and numerical computation, remains an exciting future direction.

{\it Acknowledgments --- } DC thanks E. Berg, H. Guo, S. Kivelson and S. Sachdev for useful discussions, and  H. Guo for comments on previous versions of this manuscript, which helped improve the presentation of the bound. DC is supported in part by a NSF CAREER grant (DMR-2237522), and a Sloan Research Fellowship. DC acknowledges the use of large language models (ChatGPT 5.1 and Gemini 3) for limited research assistance.
\bibliography{refs}
\clearpage
\newpage
\renewcommand{\thefigure}{S\arabic{figure}}
\renewcommand{\figurename}{Supplemental Figure}
\setcounter{figure}{0}
\renewcommand{\theequation}{S\arabic{equation}}
\setcounter{equation}{0}
\setcounter{page}{1}
\begin{widetext}
\begin{center}
    {\bf Supplementary Material for ``Planckian Bounds via Spectral Moments of Optical Conductivity"}\\
    {Debanjan Chowdhury}\\
    {Department of Physics, Cornell University, Ithaca NY 14853}
\end{center}

\section{Bounding derivatives on a complex strip}
\noindent Here we will review complex analysis to bound the derivatives of a function, $f(z)$, on the real axis, i.e. on $|f^{(n)}(t)|$, when $f(z)$ is analytic in the strip, $|\tn{Im}~z|<a$, with $|f(z)|\leq M$. Our starting point is Cauchy's integral formula on the strip,
\beq
f^{(n)}(t) = \frac{n!}{2\pi i}\oint_{C_R} \frac{f(z)}{(z-t)^{n+1}}~dz,
\eeq
where the integration is performed along a contour $C_R$ in a counterclockwise fashion. Using $\int g(z)~dz\leq \int |g(z)|~|dz|$, and noting that the maximum $R$ in the strip can be $a$, we find,
\beq
f^{(n)}(t) \leq \frac{n!~M}{a^n}.
\label{cauchy1}
\eeq

\noindent For our thermal strip of interest, $a=\beta/2$ and $M=1$. Since our interest is in the second derivative of $f(z)$ on the real axis, we have,
\beq
|f''(t)|\leq \frac{8}{\beta^2},
\eeq
i.e. $C=8$ in Eq.~\eqref{bound1}. This bound can be improved further by including the entire vertical strip (reduced to two horizontal segments, $x\pm ia$), instead of just the circle centered around $t$. It is straightforward to show that,
\beq
|f''(0)|\leq \frac{2M}{\pi}\int_{-\infty}^\infty \frac{dx}{(x^2+a^2)^{3/2}}~\bigg(=\frac{4M}{\pi a^2}\bigg),
\label{cauchystrip}
\eeq
which yields $C=16/\pi~(<8)$ for our specific problem.  

\noindent An alternative way to impose bounds on $|f^{(n)}(t)|$ is to use a combination of conformal mapping, that maps the strip to a unit disk,  and the Schwarz-Pick lemma. The conformal transformation that maps the above strip, $|\tn{Im}~z|<a$, to a unit disk, $|w|<1$ is given by $w=\tanh[\pi (z-t)/4a]$. When $z=t$, $w(t)=0$, which maps our point of interest to the origin of the disk. Then define a composed function $g(w)$ on this unit disk,
\beq
g(w)= \frac{1}{M}f(z(w)),
\eeq
where $z(w)$ is the inverse of the conformal map. Thus $g(w)$ is analytic in the disk $|w|<1$ with $|g(w)|<1$. The derivative, $f'(z) = M g'[w(z)]~w'(z)$, when applied to $z=t$ yields,
\beq
f'(t) = M~g'(0)~w'(t),\\
|f'(t)| \leq \frac{M|g'(0)|\pi}{4a}.
\eeq
For any function that maps the disk onto itself, the Schwarz-Pick lemma states,
\beq
|g'(z)|\leq \frac{1-|g(z)|^2}{1-|z|^2}.
\eeq
Thus, we have $|g'(0)|\leq 1-|g(0)|^2\leq1$. Hence,
\beq
|f'(t)| \leq \frac{M\pi}{4a}.
\eeq
For higher derivatives, this can be generalized to, 
\beq
|f^{(n)}(t)|\leq Mn!\bigg(\frac{\pi}{2a}\bigg)^n .
\eeq
For $n=2$ and $a=\beta/2$ this yields $C=2\pi^2$, which is larger than both our previous bounds.
\section{Additional details on imaginary time correlators and moments}
\noindent In Lehmann representation, the imaginary time correlator can be expressed as,
\beq
\Lambda(\tau) = \frac{1}{Z}\sum_{m,n}e^{-\beta E_m} e^{-\tau\omega_{nm}}|J_{mn}|^2,~~\tn{where}~\omega_{nm}=E_n-E_m.
\eeq
We can introduce the spectral density,
\beq
S(\omega) = \frac{2\pi}{Z} \sum_{m,n} e^{-\beta E_m}|J_{mn}|^2\delta(\omega-\omega_{nm}),
\eeq
which satisfies the standard property, $S(-\omega)=e^{-\beta\omega}S(\omega)$. The imaginary time correlator is related to the spectral density as,
\beq
\Lambda(\tau) = \int_0^\infty \frac{d\omega}{2\pi}S(\omega)2e^{-\beta\omega/2}\cosh\bigg[\bigg(\frac{\beta}{2}-\tau\bigg)\omega\bigg].\label{imT1}
\eeq
The spectral density is also related to the dissipative part of the conductivity as,
\beq
\rs = \frac{1-e^{-\beta\omega}}{2\omega}S(\omega).\label{imT2}
\eeq
Combining Eq.~\eqref{imT1}-\eqref{imT2}, we obtain,
\beq
\Lambda(\tau) = \int_{-\infty}^\infty\frac{d\omega}{2\pi}\omega\rs\frac{\cosh\bigg[\bigg(\frac{\beta}{2}-\tau\bigg)\omega\bigg]}{\sinh\bigg(\frac{\beta\omega}{2}\bigg)}.
\eeq
Finally, if we consider imaginary time derivatives of $\Lambda(\tau)$ at $\tau=\beta/2$, we find,
\beq
\partial_\tau^2\Lambda(\tau=\beta/2) &=& \int_{-\infty}^\infty\frac{d\omega}{2\pi}\frac{\omega^3\rs}{\sinh\bigg(\frac{\beta\omega}{2}\bigg)},\\
\Lambda(\tau=0) &=& \int_{-\infty}^\infty\frac{d\omega}{2\pi}\frac{\omega\rs}{\tanh\bigg(\frac{\beta\omega}{2}\bigg)},
\eeq
which are precisely related to the moments we introduced based on the Wightman correlators in the main text. 

\section{Analytical results for quasielastic Drude conductivity}
\noindent In this section, we will provide additional details on $\B$ for Drude conductivity associated with {\it quasielastic} scattering,
\beq
\rs = \frac{D\g}{\g^2 + \omega^2}. 
\eeq
The bounded quantity is of the form $\B = \M_2/\M_0$, where the numerator of interest is ,
\beq
\M_2 = \int_\omega \frac{\omega^3}{\sinh(\beta\omega/2)} ~\rs. 
\eeq
Introducing a rescaled variable, $y=\omega/\g$, and a dimensionless parameter, $\a\equiv\beta\g/2~(=\g/2T)$, we obtain,
\beq
\M_2 = D\g^3~I_2(\alpha), 
\eeq
where we introduce the integrals,
\beq
I_k(\a) &=& \int_0^\infty dy~\frac{y^{k+1}}{1+y^2}\frac{1}{\sinh(\a y)},~~k=0,2.
\eeq
The analytical forms for both can be obtained. The first one is given by,
\begin{subequations}
\beq
I_0(\a) &=& \frac{\pi}{2\sin(\a)} - \pi^2\sum_{n=1}^\infty (-1)^n\frac{n}{\a^2-(\pi n)^2},\\
I_0(\a) &=& \frac{\pi}{2\sin(\a)} - \frac{1}{4}\bigg[\psi(1-\a') + \psi(1+\a') - \psi(1/2-\a')-\psi(1/2+\a')\bigg],
\eeq
\end{subequations}
where $\psi(z)$ is the usual digamma function defined as the logarithmic derivative of the gamma function and $\a'=\a/2\pi$. The second integral is related to the first and is given by,
\begin{subequations}
\beq
I_2(\a) &=& J(\a) - I_0(\a),~\tn{where}~ J(\a)=\int_0^\infty dy~\frac{y}{\sinh(\a y)} = \frac{2}{\a^2}\sum_{n=0}^\infty\frac{1}{(2n+1)^2},\\
I_2(\a) &=& \frac{\pi^2}{4\a^2} - I_0(\a).
\eeq
\end{subequations}
It is important to note that in spite of the specific form of the results for $I_0,~I_2$, they are both well behaved for $\a=m\pi$; the seeming divergence from the $\sin(\a)$ in the denominator is {\it exactly} cancelled by the combination of digamma functions, leading to a finite answer. 

It is useful to obtain the leading behavior for the above integrals in the two asymptotic limits, $\a\ll1$, and $\a\gg1$, where we obtain,
\begin{subequations}
\beq
&&I_0(\a)\stackrel{\a\rightarrow0}{\mathlarger{\approx}} \frac{\pi}{2\a} - \ln2 + \frac{\pi\a}{12} + O(\a^2),~~I_0(\a)\stackrel{\a\gg1}{\mathlarger{\approx}}\frac{\pi^2}{4\a^2} - \frac{\pi^4}{8\a^4} + O(\a^{-6}),\\
&&I_2(\a) \stackrel{\a\rightarrow0}{\mathlarger{\approx}} \frac{\pi^2}{4\a^2} - \frac{\pi}{2\a} + \ln2 - \frac{\pi\a}{12} + O(\a^2),~~I_2(\a)\stackrel{\a\gg1}{\mathlarger{\approx}} \frac{\pi^4}{8\a^4} + O(\a^{-6}).
\eeq
\end{subequations}

\end{widetext}

\end{document}